\documentclass[a4paper,12pt,reqno]{amsart}

\usepackage{graphicx,amssymb,datetime,float,MnSymbol,currfile,tikz,hyperref,multirow}
\usepackage[square,comma,numbers]{natbib}
\usepackage[foot]{amsaddr}
\usetikzlibrary{arrows}
\usetikzlibrary{decorations.markings}

\def\ci{\!\perp\!}

\newcommand{\comments}[1]{}

\DeclareMathOperator*{\argmin}{arg\,min}

\tikzset{tt/.style={decoration={
  markings,
  mark=at position .485 with {\arrow{>}},
  mark=at position .515 with {\arrow{<}}},postaction={decorate}}}

\begin{document}

\title[]{Bounds and Sensitivity Analysis of the Causal Effect Under Outcome-Independent MNAR Confounding}

\author{Jose M. Pe\~{n}a$^1$}
\address{$^1$Link\"oping University, Sweden.}
\email{jose.m.pena@liu.se}


\begin{abstract}
We report assumption-free bounds for any contrast between the probabilities of the potential outcome under exposure and non-exposure when the confounders are missing not at random. We assume that the missingness mechanism is outcome-independent. We also report a sensitivity analysis method to complement our bounds.
\end{abstract}

\maketitle

\begin{figure}[t]
\begin{tikzpicture}[inner sep=1mm]
\node at (0,0) (E) {$E$};
\node at (4,0) (D) {$D$};
\node at (2,3) (U) {$U$};
\node at (2,1) (R) {$R$};
\path[->] (E) edge (D);
\path[->] (U) edge (E);
\path[->] (U) edge (D);
\path[->] (U) edge[dashed] (R);
\path[->] (E) edge[dashed] (R);
\end{tikzpicture}
\caption{Causal graph where the dashed edges represent the missingness mechanism.}\label{fig:graph}
\end{figure}
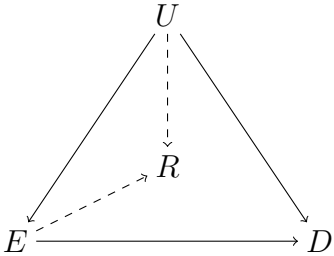

\section{Introduction}\label{sec:introduction}

This article is concerned with causal effect estimation in the graph in Figure \ref{fig:graph}, where $E$ denotes the exposure, $D$ denotes the outcome, and $U$ denotes the confounder. Moreover, $E$ and $D$ are binary, and $U$ is categorical. While $E$ and $D$ are fully observed, $U$ is partially observed. The binary random variable $R$ indicates whether the value of $U$ is observed ($R=0$) or missing. In the causal graph in Figure \ref{fig:graph}, the dashed edges represent the missingness mechanism. 

The work \cite{MohanandPearl2021} classifies missingness mechanisms into three types:
\begin{itemize}
    \item missing completely at random (MCAR) if $R \ci D,E,U$, and
    \item missing at random (MAR) if $R \ci U | D,E$, and
    \item missing not at random (MNAR) otherwise.
\end{itemize}
Therefore, the missingness mechanism in the graph in Figure \ref{fig:graph} is MNAR. More specifically, it is outcome-independent MNAR because $R \ci D | E,U$. This mechanism is plausible if $R$ is measured before $D$. Previous works studying such a mechanism include \cite{DingandGeng2014,Yangetal.2019}.

Two of the most common approaches to deal with missing data are arguably complete case analysis and multiple imputation \cite{Rubin1987}. The former simply drops all the cases with missing values and analyzes the remaining cases. The latter replaces each missing value by multiple values drawn from the posterior distribution of the missing value given the complete data, so as to account for the uncertainty about the missing value. The imputed values are used to create completed datasets that are analyzed as if there had been no missing data. The results obtained from the completed datasets are then combined. Complete case analysis gives unbiased estimates only under the MCAR mechanism, whereas multiple imputation gives unbiased estimates only under the MCAR and MAR mechanisms. Therefore, neither method should be expected to recover the risk ratio (or any other contrast) from the observed probability distributions $\{p(D,E,U,R=0), p(D,E,R=1)\}$ for the causal graph in Figure \ref{fig:graph}. However, the risk ratio can be recovered for this graph by solving a system of linear equations if $U$ does not have more categories than $D$ \cite{DingandGeng2014,Yangetal.2019}. Therefore, this method is not suitable for our problem either, since $U$ can have any number of categories.

Since in general point estimation is not possible under MNAR confounding, sensitivity analysis methods have been developed. The work \cite{EglestonandWong2009} presents a method for sensitivity analysis under MNAR confounding for survival analysis, which thus is not suitable for our goal of estimating the risk ratio (or any other contrast). Moreover, the authors assume a particular model for the missingness mechanism, which we do not. The work \cite{LuandAshmead2018} presents another method for sensitivity analysis under MNAR confounding. The method bounds the $p$-value of a test of no causal effect as a function of a sensitivity parameter, which is different from our aim of estimating a contrast. 

Finally, it is worth mentioning some works that address problems close to ours. The work \cite{HorowitzandManski2000} derives sharp bounds for the risk difference conditioned on the state of the confounders under MNAR confounding. How to combine these bounds to produce sharp bounds of the risk difference is an open problem. The work \cite{MohanandPearl2021} gives examples where the risk ratio is recoverable under MNAR for some variables but the confounders, which are fully observed. The work \cite{SunandFu2023} establishes conditions for recoverability of the risk difference under treatment-independent MNAR confounding, i.e., $R \ci E | D,U$. This missingness mechanism is also considered in the work \cite{MiaoandTchetgenTchetgen2018} proving parameter identification in some semiparametric models.

The rest of this article is organized as follows. We derive assumption-free bounds for any contrast in Section \ref{sec:bounds}, and compare them to complete case analysis and multiple imputation with an example and simulated experiments in Sections \ref{sec:example} and \ref{sec:experiments}. We complement our bounds by deriving a sensitivity analysis method in Section \ref{sec:SA}, which we illustrate with an example and simulated experiments in Sections \ref{sec:example2} and \ref{sec:experiments2}. We close with a brief discussion in Section \ref{sec:discussion}. The R code for our examples and experiments is available \href{https://www.dropbox.com/scl/fi/9qwg5758jgol2y6mvx9z3/MNAR5.R?rlkey=kfqaqrgcqapo7vs6rg318zxjd&st=appkqg1t&dl=0}{here}.

\section{Bounds}\label{sec:bounds}

The causal graph in Figure \ref{fig:graph} represents a non-parametric structural equation model with independent errors, which defines a joint probability distribution $p(D,E,U,R)$. We make the usual assumption that $p(D,E,U,R=0)>0$ so that recoverability of a quantity from $p(D,E,U,R=0)$ can be judged from the graph \cite{MohanandPearl2021}. We use upper-case letters to denote random variables, and the same letters in lower-case to denote their values. 

Let $D_e$ denote the potential outcome when the exposure is set to level $E=e$. Note that
\begin{align}\label{eq:D1}\nonumber
p(D_e=1) & = p(D_e=1 | E=e) p(E=e) + p(D_e=1 | E=1-e) p(E=1-e)\\
& = p(D=1 | E=e) p(E=e) + p(D_e=1 | E=1-e) p(E=1-e)
\end{align}
where the second equality follows from counterfactual consistency, i.e., $E=e \Rightarrow D_e = D$. Moreover,
\begin{align*}
p(D_e=1 | E=1-e) & = p(D_e=1 | E=1-e, R=0) p(R=0|E=1-e)\\
& + p(D_e=1 | E=1-e, R=1) p(R=1|E=1-e)
\end{align*}
where
\begin{align*}
& p(D_e=1 | E=1-e, R=0)\\
& = \sum_u p(D=1 | E=e, U=u, R=0) p(U=u | E=1-e, R=0)
\end{align*}
due to $D_e \ci E | U, R$ for all $e$ in the causal graph under consideration, and counterfactual consistency.\footnote{In the previous equation, we could drop $R=0$ from the conditioning set since $D \ci R | E, U$. However, we prefer to keep it to explicitly indicate that $p(D|E,U)$ can be recovered from the observed probability distribution $p(D,E,U,R=0)$ as $p(D|E,U,R=0)$.} We next bound $p(D_e=1 | E=1-e, R=1)$ in terms of recoverable quantities. Specifically,
\begin{align}\nonumber
& p(D_e=1 | E=1-e, R=1)\\\nonumber
& = \sum_u p(D=1 | E=e, U=u, R=1) p(U=u | E=1-e, R=1)\\\label{eq:aux}
& = \sum_u p(D=1 | E=e, U=u, R=0) p(U=u | E=1-e, R=1)
\end{align}
where the first equality follows from $D_e \ci E | U, R$ for all $e$ in the causal graph under consideration, and counterfactual consistency, and the second equality follows from $D \ci R | E,U$ in the causal graph under consideration (i.e., outcome-independent MNAR confounding). Since $p(U|E,R=1)$ is not recoverable,\footnote{Otherwise $p(U|E)$ would be recoverable as $p(U|E,R=0) p(R=0|E) + p(U|E,R=1) p(R=1|E)$, which is not \cite[Theorem 3]{MohanandPearl2021}.} we resort to bounding $p(D_e=1 | E=1-e, R=1)$ as
\[
m(e) \leq p(D_e=1 | E=1-e, R=1) \leq M(e)
\]
where
\[
m(e)=\min_{u} p(D=1 | E=e, U=u, R=0).
\]
and
\[
M(e)=\max_{u} p(D=1 | E=e, U=u, R=0).
\]
Incorporating these bounds in Equation \ref{eq:D1} gives the following bounds of $p(D_e=1)$:
{\footnotesize
\begin{align}\label{eq:D1bound}\nonumber
& p(D=1, E=e) + p(D_e=1 | E=1-e, R=0) p(R=0, E=1-e) + p(R=1, E=1-e) m(e)\\
& \leq p(D_e=1) \leq\\\nonumber
& \min \big( 1, p(D=1, E=e) + p(D_e=1 | E=1-e, R=0) p(R=0, E=1-e) + p(R=1, E=1-e) M(e) \big).
\end{align}
}

Finally, we can obtain a lower (resp. upper) bound for any contrast between $p(D_1=1)$ and $p(D_0=1)$ by contrasting the lower (resp. upper) bound for $p(D_1=1)$ and the upper (resp. lower) bound for $p(D_0=1)$ in the previous equation. For instance, we can obtain bounds for the risk ratio, risk difference, odds ratio, odds difference, etc. Note that the bounds are assumption-free. However, they are not sharp (see the appendix).

\subsection{Example}\label{sec:example}

Consider the following distribution $p(D,E,U,R)$ compatible with the causal graph in Figure \ref{fig:graph} and where $U$ is ternary:
\begin{align*}
&p(U)=(0.4,0.5,0.1)\\
&p(E=1|U)=(0.3,0.1,0.2)\\
&p(D=1|E=0,U)=(0.1,0.9,0.7)\\
&p(D=1|E=1,U)=(0.8,0.3,0.2)\\
&p(R=1|E=0,U)=(0.1,0.95,0.85)\\
&p(R=1|E=1,U)=(0.2,0.8,0.9).
\end{align*}

Let $U$ represent the income of the individuals in the population. The treatment has a positive effect on low income individuals ($U=0$), but it has a negative effect on medium and high income individuals. Low income individuals are more likely to report their income status than medium and high income individuals. 

Since we have access to the data generation process, we can compute the true risk ratio $RR_{true}=p(D_1=1)/p(D_0=1)$ by adjusting for $U$, that is
\begin{equation}\label{eq:RR}
 p(D_e=1)=\sum_u p(D=1|E=e,U=u) p(U=u).   
\end{equation}
In reality, when the data generation process is unknown, we can perform a complete case analysis, which in our case amounts to adjusting for $U|R=0$, that is replacing $p(U)$ in Equation \ref{eq:RR} with $p(U|R=0)$. We refer to this quantity as $RR_{CC}$. We can also apply multiple imputation to obtain a quantity that we refer to as $RR_{MI}$. In our case, this amounts to replacing $p(U)$ in Equation \ref{eq:RR} with
\begin{align}\nonumber
& p(U|R=0) p(R=0)\\\label{eq:MI}
& + p(R=1) \sum_{d,e} p(U|D=d,E=e,R=0) p(D=d,E=e|R=1).   
\end{align}
We explain next the correspondence between the terms in this equation and the steps in multiple imputation applied to an incomplete dataset (recall Section \ref{sec:introduction}). The first term in the sum corresponds to the distribution of $U$ derived from the complete cases, whereas the second term in the sum corresponds to the distribution of $U$ derived from the cases completed through imputation. The term $p(U|D=d,E=e,R=0)$ represents the uncertainty in the value to impute for the incomplete cases where $D=d$ and $E=e$, which is assessed from the complete cases. The term $p(D=d,E=e|R=1)$ represents the fraction of incomplete cases where $D=d$ and $E=e$. Finally, our alternative to complete case analysis and multiple imputation is to lower and upper bound $RR_{true}$, respectively, by
\[
LB=LB(1)/UB(0)
\]
and
\[
UB=UB(1)/LB(0)
\]
where $LB(e)$ and $UB(e)$ denote the lower and upper bounds of $p(D_e=1)$ in Equation \ref{eq:D1bound}.

In this example, $RR_{true}=0.88$, $RR_{CC}=3.94$ and $RR_{MI}=2.01$. Our bounds constrain $RR_{true}$ to lie in the interval $[0.74, 1.44]$. Therefore, $RR_{CC}$ and $RR_{MI}$ are so biased that they wrongly suggest a positive effect of the treatment on the whole population. Moreover, they are so biased that they lie outside our interval, and thus they are not logically possible. The reason of this is that the majority of the complete cases correspond to the subpopulation $U=0$ (i.e., low income individuals), who benefit from the treatment. Specifically, compare $p(U=0)=0.4$ with $p(U=0|R=0)=0.88$ and 0.7 for Equation \ref{eq:MI} with $U=0$.

Note that our interval includes the null causal effect. This is a consequence of the non-recoverability of the association between $E$ and $U$, which may be so strong as to nullify the causal effect or even reverse it. Note though that our interval does not need to be centered at the null causal effect, and thus it may be informative about both the magnitude and log-sign of the true causal effect. In any case, recall that our bounds are assumption-free.

\subsection{Experiments}\label{sec:experiments}

We report some experiments that provide additional evidence on the observations made with the previous example. Specifically, we generate $10^6$ distributions $p(D,E,U,R)$ compatible with the causal graph in Figure \ref{fig:graph} and where $U$ is ternary. All the parameters are sampled uniformly from the interval $[0,1]$. This corresponds to a MNAR mechanism. We count the number of generated distributions where (i) $RR_{CC}$ and $RR_{MI}$ disagree with $RR_{true}$ (i.e., the former are biased), (ii) they disagree on the log-sign (i.e., they disagree on whether the treatment is beneficial or not), (iii) $RR_{CC}$ and $RR_{MI}$ are outside the bounds derived in Section \ref{sec:bounds}, and (iv) both (ii) and (iii) occur. We repeat the experiments after removing the edge $U \dashrightarrow R$, which corresponds to a MAR mechanism. Finally, we do the same removing both incoming edges to $R$, which corresponds to a MCAR mechanism.

The first six rows of Table \ref{tab:results1} present the results (in percentages). These follow the expected pattern discussed in Section \ref{sec:introduction}, namely $RR_{CC}$ and $RR_{MI}$ are unbiased under the MCAR mechanism, $RR_{CC}$ is biased and $RR_{MI}$ unbiased under the MAR mechanism, and both $RR_{CC}$ and $RR_{MI}$ are biased under the MNAR mechanism.\footnote{In these experiments, complete case analysis can be made unbiased for the MAR mechanism by first recovering $p(D,E,U)$, which is recoverable for that mechanism in the causal graph under consideration \cite[Theorem 1]{MohanandPearl2021}.} For the MNAR mechanism, moreover, a non-negligible percentage of $RR_{CC}$ and $RR_{MI}$ have incorrect log-sign and/or lie outside our bounds. Some particular missingness patterns may exacerbate these problems. For instance, the last two rows of Table \ref{tab:results1} present the results of repeating our experiments but keeping $p(R|E,U)$ fixed to the values in Section \ref{sec:example}. Therefore, our bounds are a safe alternative to $RR_{CC}$ and $RR_{MI}$, as they always include $RR_{true}$. Moreover, recall that they are assumption-free.

\begin{table}[t]
    \caption{Results of the experiments in Section \ref{sec:experiments} (in percentages).}
    \label{tab:results1}
    \centering
    \begin{tabular}{l|l|r|r|r|r}
                Missingness & Method & Biased & Wrong log-sign & Out bounds & Both\\
                \hline
        MCAR    & CC &  0   & 0.0 & 0.0 & 0.0\\
                & MI &  0   & 0.0 & 0.0 & 0.0\\
        MAR     & CC &  100 & 4.3   & 5.9   & 0.7\\
                & MI &  0 &   0.0 &   0.0 & 0.0\\
        MNAR    & CC &  100 & 9.3   & 15.4   & 2.6\\
                & MI &  100 & 7.5   & 8.9   & 1.2\\
                \hline
        MNARex    & CC &  100 & 20.7   & 41.4   & 11.6\\
                & MI &  100 & 16.1   & 25.8   & 5.3\\
    \end{tabular}
\end{table}

Our experiments provide additional evidence on the bias induced by complete case analysis and multiple imputation under MNAR confounding, which is related to the bias induced by adjusting for a proxy of an unmeasured confounder \cite{Gabrieletal.2022}.

\section{Sensitivity Analysis}\label{sec:SA}

Unlike in the previous section, we now bound any contrast of interest in terms of some user-defined parameters. Recall that $p(U|E,R=1)$ is not recoverable. Let us define the sensitivity parameters
\[
\alpha(e)=\min_{u} p(U=u|E=e,R=1)
\]
and
\[
\beta(e)=\max_{u} p(U=u|E=e,R=1)
\]
whose values the analyst has to specify for all $e$. By definition, these values must lie in the interval $[0,1]$ and $\alpha(e) \leq \beta(e)$. The observed data distributions constrain the valid values further. To see it, note that
\begin{align*}
p(D=1|E=e,R=1) &= \sum_u p(D=1|E=e,U=u,R=1) p(U=u|E=e,R=1)\\
&= \sum_u p(D=1|E=e,U=u,R=0) p(U=u|E=e,R=1)\\
&\geq \alpha(e) \sum_u p(D=1|E=e,U=u,R=0)   
\end{align*}
where the second equality follows from $R \ci D | E,U$, and thus
\[
\alpha(e) \leq \min \big( 1, p(D=1|E=e,R=1) / \sum_u p(D=1|E=e,U=u,R=0) \big)
\]
and likewise
\[
\beta(e) \geq \min \big( 1, p(D=1|E=e,R=1) / \sum_u p(D=1|E=e,U=u,R=0) \big).
\]
We can thus define the feasible regions for $\alpha(e)$ and $\beta(e)$ as
\[
0 \leq \alpha(e) \leq \min \big( 1, p(D=1|E=e,R=1) / \sum_u p(D=1|E=e,U=u,R=0) \big) \leq \beta(e) \leq 1.
\]

Now, note that Equation \ref{eq:aux} implies that
\begin{align}\label{eq:aux2}\nonumber
&\alpha(1-e) \sum_u p(D=1 | E=e, U=u, R=0)\\
&\leq p(D_e=1 | E=1-e, R=1) \leq\\\nonumber
&\beta(1-e) \sum_u p(D=1 | E=e, U=u, R=0).
\end{align}
Therefore, as before, we can incorporate these bounds in Equation \ref{eq:D1} to bound $p(D_e=1)$, which allows us to obtain a lower (resp. upper) bound for any contrast between $p(D_1=1)$ and $p(D_0=1)$ by contrasting the lower (resp. upper) bound for $p(D_1=1)$ and the upper (resp. lower) bound for $p(D_0=1)$. However, the bounds are not sharp (see the appendix). To ensure that the bounds include the true value of the contrast, the analyst has to choose values for $\alpha(e)$ and $\beta(e)$ such that Equation \ref{eq:aux2} holds. In other words, the analyst has to choose values for $\alpha(e)$ and $\beta(e)$ such that
\[
\alpha(e) \leq \min_{u} p(U=u|E=e,R=1)
\]
and
\[
\beta(e) \geq \min_{u} p(U=u|E=e,R=1)
\]
but she can never be certain of this, since $p(U|E,R=1)$ is not recoverable. To put it differently, some values in the feasible regions of $\alpha(e)$ and $\beta(e)$ may result in bounds that do not contain the true value of the contrast. Even if they do contain it, there is no guarantee that these bounds are narrower than the assumption-free bounds derived in Section \ref{sec:bounds}. Finally, the fact that each bound only involves two sensitivity parameters (i.e., $\alpha(0)$ and $\beta(1)$ for the lower bound, and $\alpha(1)$ and $\beta(0)$ for the upper bound) makes the sensitivity analysis easy to visualize in tables or 2-D plots.\footnote{If the analyst considers that having to specify four sensitivity parameters is too demanding, then one can derive a similar sensitivity analysis method with just two parameters, namely $\alpha=\min_{e,u} p(U=u|E=e,R=1)$ and $\beta=\max_{e,u} p(U=u|E=e,R=1)$.} We illustrate these observations in the next two sections.

\begin{figure}[t]
\centering
\includegraphics[scale=.75]{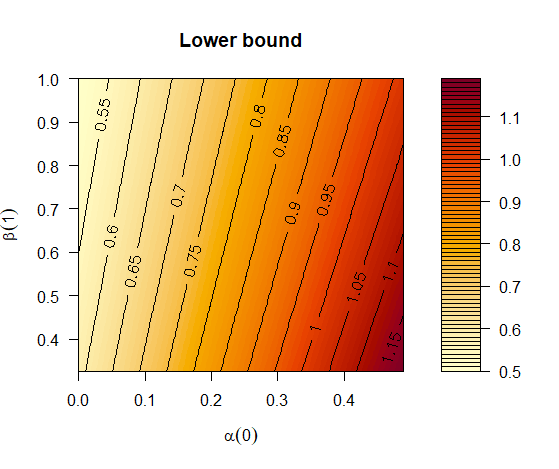}
\includegraphics[scale=.75]{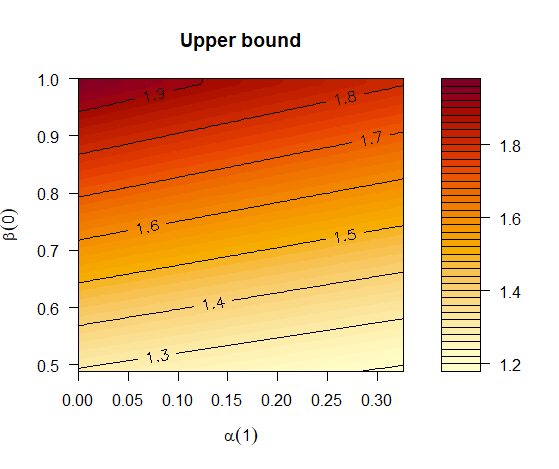}
\caption{Bounds for the example in Section \ref{sec:example2}.}\label{fig:example2}
\end{figure}

\subsection{Example}\label{sec:example2}

We illustrate our sensitivity analysis method with the example in Section \ref{sec:example}. Figure \ref{fig:example2} shows the bounds of $RR_{true}$ as a function of the sensitivity parameters. The axes span the feasible regions of the parameters. In the data generation model considered, $\alpha(0)=0.05$, $\alpha(1)=0.22$, $\beta(0)=0.82$ and $\beta(1)=0.49$. These values are unknown to the analyst, because $p(U|E,R=1)$ is not recoverable. However, the figure reveals that the analyst only needs to have some rough idea of these values to confidently conclude that $RR_{true}$ lies in the interval $[0.6,1.7]$. Note that this interval is wider than the assumption-free interval reported in Section \ref{sec:example}. We investigate this question further in the next section. Narrower intervals can be obtained at the risk of excluding $RR_{true}$. For instance, choosing $\alpha(0)=\beta(1)=0.4$ results in a lower bound of 1.05, whereas $RR_{true}=0.88$. Therefore, the analyst should act conservatively and choose values that she believes are not greater (resp. smaller) than the unknown true values of $\alpha(e)$ (resp. $\beta(e)$).

\subsection{Experiments}\label{sec:experiments2}

We report some experiments showing that, unlike in the previous section, our sensitivity analysis method quite often produces bounds that are narrower than those derived in Section \ref{sec:bounds}, while including $RR_{true}$. Specifically, we generate $10^6$ distributions $p(D,E,U,R)$ compatible with the causal graph in Figure \ref{fig:graph} and where $U$ is ternary. All the parameters are sampled uniformly from the interval $[0,1]$. This corresponds to a MNAR mechanism. For each distribution generated, we compute the true values of the sensitivity parameters, which we denote by $\alpha^*(e)$ and $\beta^*(e)$. These are unknown to the analyst. Instead, the analyst chooses values for the sensitivity parameters as
\[
\alpha(e)=\min(1,\alpha^*(e)/f)
\]
and
\[
\beta(e)=\min(1,\beta^*(e) \cdot f)
\]
with factor $f=0.9, 1, 1.1, 1.2$. This amounts to considering four types of analysts, from slightly risky (i.e., $f=0.9$) to optimal (i.e., $f=1$) to conservative (i.e., $f=1.1,1.2$). For each analyst type, we count the number of generated distributions where (i) the sensitivity analysis bounds include $RR_{true}$, (ii) the sensitivity analysis bounds are narrower than the assumption-free bounds derived in Section \ref{sec:bounds}.

Table \ref{tab:results2} presents the results (in percentages). As expected, the larger the factor $f$ the wider the sensitivity analysis interval, and thus the less likely to improve the assumption-free bounds but the more likely to include $RR_{true}$. Being slightly risky (i.e., $f=0.9$) improves the assumption-free bounds without missing $RR_{true}$ in most cases. Being conservative (i.e., $f=1.1,1.2$) improves the assumption-free bounds in a sizeable number of cases without never missing $RR_{true}$. In summary, if the analyst is able to produce relatively accurate values for the sensitivity parameters, then there is a good chance that our sensitivity analysis method returns bounds that include $RR_{true}$ and are narrower than the assumption-free ones. This calls for combining both pairs of bounds by taking the narrowest ones.

\begin{table}[t]
    \caption{Results of the experiments in Section \ref{sec:experiments2} (in percentages).}
    \label{tab:results2}
    \centering
    \begin{tabular}{l|r|r|r|r}
                Factor & $RR_{true}$ included & LB narrower & UB narrower & Both narrower\\
                \hline
        0.9     &  99.7 & 26.6   & 26.6  & 16.5\\
        1    &  100   & 18.9 & 18.9 & 10.7\\

        1.1    &  100 & 13.2   & 13.2   & 6.7\\
        1.2    &  100 & 9.3   & 9.2   & 4.1\\

    \end{tabular}
\end{table}

\section{Discussion}\label{sec:discussion}

We have derived assumption-free bounds for any contrast between the probabilities of the potential outcome under exposure and non-exposure when there is outcome-independent MNAR confounding. Our bounds are therefore a safe alternative to commonly used solutions such as complete case analysis and multiple imputation which can be very biased, as shown by our example and experiments. We have complemented our bounds with a sensitivity analysis method to produce narrower bounds.

In the future, we would like to derive similar bounds and/or a sensitivity analysis method for standard MNAR confounding. An extension of our work that we are currently studying is the case where the analyst does not specify the values of the sensitivity parameters but their distribution. We can then use Mote Carlo simulation to approximate the distributions of the lower and upper bounds for any contrast: Sample values for the sensitivity parameters, compute the bounds, and repeat. We can finally use the approximated distributions to approximate the expectations of the bounds or the probabilities that the bounds are smaller or bigger than a given value.

Finally, it is worth mentioning that the problem considered in this article has a straightforward solution via the data fusion approach proposed in the work \cite{Wangetal.2024}. This approach consists in augmenting the primary MNAR dataset with some hopefully available MAR dataset. In our problem, the primary dataset is over $\{D,E,U\}$ where $U$ is MNAR and thus $p(D|E,U)$ can be estimated as $p(D|E,U,R=0)$, and the auxiliary dataset should be over $\{E,U\}$ where $U$ is MAR and thus $p(U|E)$ can be estimated as $p(E,U|R=0)/p(E|R=0)$ and used in the place of $p(U|E,R=1)$ for the primary dataset. Plugging these estimates into Equation \ref{eq:aux} gives the desired answer.

\section*{Appendix: On the Sharpness of the Bounds}

We show below that the assumption-free bounds derived in Section \ref{sec:bounds} are not sharp. Let the set $\{p'(D,E,U,R=0),p'(D,E,R=1)\}$ represent the observed data distributions at hand. To show that the lower bound for $p(D_1=1)$ in Equation \ref{eq:D1bound} is sharp, we need to construct a distribution $p(D,E,U,R)$ such that (i) $\{p(D,E,U,R=0),p(D,E,R=1)\}$ coincides with $\{p'(D,E,U,R=0),p'(D,E,R=1)\}$, and (ii) the lower bound for $p(D_1=1)$ coincides with $p(D_1=1)$. To this end, set $p(R)=p'(R)$, $p(D,E,U|R=0)=p'(D,E,U|R=0)$, and $p(E|R=1)=p'(E|R=1)$. Note that condition (ii) requires that $p(D_1=1 | E=0,R=1)=m(1)$, which implies by Equation \ref{eq:aux} that we need to set $p(U=u^* | E=0,R=1)=1$ if and only if $u^*=\argmin_{u} p'(D=1 | E=1, U=u, R=0)$. However, this may invalidate the following equality, which is required by condition (i):
\begin{align*}
p(D=1|E=0,R=1) &= \sum_u p(D=1|E=0,U=u,R=1) p(U=u|E=0,R=1)\\
&= \sum_u p(D=1|E=0,U=u,R=0) p(U=u|E=0,R=1)
\end{align*}
where the second equality follows from $R \ci D | E,U$. Therefore, the lower bound for $p(D_1=1)$ in Equation \ref{eq:D1bound} is not sharp.

Similarly, we show below that the sensitivity analysis bounds derived in Section \ref{sec:SA} are not sharp. Let the set $\{p'(D,E,U,R=0),p'(D,E,R=1)\}$ represent the observed data distributions at hand, and consider $\alpha(0)=\alpha(1)=0$ and $\beta(0)=\beta(1)=1$. Note that these values belong to the feasibility regions. To show that the lower bound for $p(D_1=1)$ is sharp, we need to construct a distribution $p(D,E,U,R)$ such that (i) $\{p(D,E,U,R=0),p(D,E,R=1)\}$ coincides with $\{p'(D,E,U,R=0),p'(D,E,R=1)\}$, (ii) the true values of the sensitivity parameters for $p(D,E,U,R)$ coincide with $\{\alpha(0), \alpha(1), \beta(0), \beta(1)\}$, and (iii) the lower bound for $p(D_1=1)$ coincides with $p(D_1=1)$. To this end, set $p(R)=p'(R)$, $p(D,E,U|R=0)=p'(D,E,U|R=0)$, and $p(E|R=1)=p'(E|R=1)$. Note that condition (iii) requires that $p(D_1=1 | E=0,R=1)=0$ by Equation \ref{eq:aux2}, which implies by Equation \ref{eq:aux} that we need to set $p(U=u | E=0,R=1)=0$ for all $u$, because we assumed in Section \ref{sec:bounds} that $p(D,E,U,R=0)>0$. This is not possible, and thus the lower bound for $p(D_1=1)$ is not sharp.

\bibliographystyle{unsrt}
\bibliography{sensitivityAnalysis}

\begin{thebibliography}{10}

\bibitem{MohanandPearl2021}
K.~Mohan and J.~Pearl.
\newblock {Graphical Models for Processing Missing Data}.
\newblock {\em Journal of the American Statistical Association}, pages 1023--1037, 2021.

\bibitem{DingandGeng2014}
P.~Ding and Z.~Geng.
\newblock {Identifiability of Subgroup Causal Effects in Randomized Experiments with Nonignorable Missing Covariates}.
\newblock {\em Statistics in Medicine}, 33:1121--1133, 2014.

\bibitem{Yangetal.2019}
S.~Yang, L.~Wang, and P.~Ding.
\newblock {Causal Inference with Confounders Missing not at Random}.
\newblock {\em Biometrika}, 106:875--888, 2019.

\bibitem{Rubin1987}
D.~B. Rubin.
\newblock {\em {Multiple Imputation for Nonresponse in Surveys}}.
\newblock John Wiley \& Sons, 1987.

\bibitem{EglestonandWong2009}
B.~L. Egleston and Y.~N. Wong.
\newblock {Sensitivity Analysis to Investigate the Impact of a Missing Covariate on Survival Analyses Using Cancer Registry Data}.
\newblock {\em Statistics in Medicine}, 28:1498--1511, 2009.

\bibitem{LuandAshmead2018}
B.~Lu and R.~Ashmead.
\newblock {Propensity Score Matching Analysis for Causal Effects with MNAR Covariates}.
\newblock {\em Statistica Sinica}, 28:2005--2025, 2018.

\bibitem{HorowitzandManski2000}
J.~L. Horowitz and C.~F. Manski.
\newblock {Nonparametric Analysis of Randomized Experiments with Missing Covariate and Outcome Data}.
\newblock {\em Journal of the American Statistical Association}, 95:77--84, 2020.

\bibitem{SunandFu2023}
J.~Sun and B.~Fu.
\newblock {Identification and Estimation of Causal Effects with Confounders Missing Not at Random}.
\newblock {\em arXiv:2303.05878 [stat.ME]}, 2023.

\bibitem{MiaoandTchetgenTchetgen2018}
W.~Miao and E.~Tchetgen~Tchetgen.
\newblock {Identification and Inference with Nonignorable Missing Covariate Data}.
\newblock {\em Statistica Sinica}, 28:2049--2067, 2018.

\bibitem{Gabrieletal.2022}
E.~E. Gabriel, J.~M. Pe\~{n}a, and A.~Sjölander.
\newblock {Bias Attenuation Results for Dichotomization of a Continuous Confounder}.
\newblock {\em Journal of Causal Inference}, 10:515--526, 2022.

\bibitem{Wangetal.2024}
Z.~Wang, A.~Ghassami, and I.~Shpitser.
\newblock {Identification and Estimation for Nonignorable Missing Data: A Data Fusion Approach}.
\newblock In {\em Proceedings of the 41st International Conference on Machine Learning}, pages 50467--50488, 2024.

\end{thebibliography}

\end{document}